\documentclass[aps,prd,twocolumn,superscriptaddress]{revtex4}
\usepackage{epsfig,epsf}
\usepackage{amsmath}
\usepackage{amsthm}
\usepackage{amsfonts}
\usepackage{amssymb}
\usepackage{dsfont}
\usepackage{multirow}
\usepackage{appendix}
\usepackage{slashed}
\usepackage[active]{srcltx}
\usepackage{psfrag}

\begin{document}

\title{Heavy double-gluon quarkonium and $\overline{b}ggc$ mesons }
\date{\today}
\author{G.~Daylan Esmer}
\affiliation{Department of Physics, Istanbul University, Vezneciler, 34134 Istanbul, T%
\"{u}rkiye}
\author{B.~Barsbay}
\affiliation{Division of Optometry, School of Medical Services and Techniques,Dogus University, Dudullu-\"{U}mraniye, 34775
Istanbul, T\"{u}rkiye}
\author{K.~Azizi}
\thanks{Corresponding author: kazizi@dogus.edu.tr}
\affiliation{Department of Physics, Dogus University, Dudullu-\"{U}mraniye, 34775
Istanbul, T\"{u}rkiye}
\author{H.~Sundu}
\affiliation{Department of Physics Engineering, Istanbul Medeniyet University, 34700
Istanbul, T\"{u}rkiye}
\author{S.~T\"{u}rkmen}
\affiliation{Department of Physics, Istanbul University, Vezneciler, 34134 Istanbul, T%
\"{u}rkiye}

\begin{abstract}
We investigate heavy double-gluon hybrid mesons, built from a heavy quark-antiquark pair and two valence gluons, using the two-point QCD sum rule method. The charmonium-like $\bar{c}GGc$ and bottomonium-like $\bar{b}GGb$ states are considered with $J^{PC}=0^{\pm\pm}$ and
$1^{\pm\pm}$, whereas the open-flavor $\bar{b}GGc$ states are classified by $J^{P}=0^{\pm}$ and $1^{\pm}$. The currents with $J^{PC}=0^{+-}, 0^{--}, 1^{++}$, and $1^{-+}$ vanish identically, so sum rules are derived for the remaining four channels. The operator product expansion is carried out up to dimension twelve, and the Borel parameter and continuum threshold are fixed so that the pole contribution dominates and the OPE converges. The resulting masses lie in the ranges $M\simeq 6.2$--$6.7$~GeV for $\bar{c}GGc$, $M\simeq 12.0$--$12.2$~GeV for $\bar{b}GGb$, and $M\simeq 9.2$--$9.6$~GeV for $\bar{b}GGc$, and the current couplings are obtained for all channels. These predictions provide a basis for analyzing the strong decays of double-gluon hybrids and for interpreting future experimental data in the heavy-quark sector.
\end{abstract}

\maketitle


\section{Introduction}

\label{sec:Intro}

Hadrons are composite particles formed through the binding of quarks and gluons under the strong interaction, which is described by Quantum Chromodynamics (QCD), one of the fundamental theoretical components of the Standard Model~\cite{Gross:1973id,Politzer:1973fx,Wilson:1974sk,Gell-Mann:1964ewy,Zweig:1964ruk}. According to the conventional quark model, all established hadrons are classified into two main categories according to their quark content: baryons and mesons. However, QCD also allows the existence of more complicated hadronic configurations beyond the conventional quark model. In recent years, theoretical predictions of such unconventional states, composed of different quark and gluon combinations, have attracted considerable attention, and some of them have been supported experimentally. These nonconventional hadronic configurations are generally referred to as exotic hadrons, including tetraquarks, pentaquarks, hybrid hadrons, meson molecules, dibaryons, and glueballs.

Over the past two decades, remarkable progress has been achieved in the study of multiquark systems owing to advances in experimental techniques and the high-precision data provided by modern high-energy collider facilities. In particular, a series of charmonium- and bottomonium-like $XYZ$ states have been observed experimentally~\cite{Belle:2003nnu,BaBar:2005hhc,Belle:2011aa,BESIII:2013ris,Belle:2013yex}. Moreover, hidden-charm pentaquark states, namely the $P_c$ states, have also been reported by the LHCb Collaboration~\cite{LHCb:2015yax,LHCb:2019kea,LHCb:2021chn}. For comprehensive reviews on exotic hadron spectroscopy, see Refs.~\cite{Chen:2016qju,Guo:2017jvc,Li:2022fgd,Brambilla:2019esw}. The recent experimental observation of new exotic candidates has further stimulated theoretical efforts toward understanding their internal structure and underlying dynamics.

Among exotic hadrons, hybrid states differ from conventional mesons and baryons in that they generally contain one or more valence gluons in addition to their valence quarks. The simplest hybrid meson configuration is the quark-antiquark-gluon state, $\bar{q}gq$, which is commonly referred to as a single-gluon hybrid. In such systems, the gluon does not merely play the role of a binding mediator between quarks, but also acts as an active valence degree of freedom that contributes directly to the quantum numbers and internal dynamics of the hadron. As a consequence, hybrid mesons may carry exotic quantum numbers that are not accessible within the conventional quark-antiquark picture, making them particularly attractive candidates for experimental identification. Recent high-precision experiments have provided increasingly strong indications for the possible existence of such states in nature.

Double-gluon hybrid states are hadronic systems in which, in addition to a quark-antiquark pair, two gluons appear as valence constituents. In such configurations, gluons do not merely act as binding mediators between quarks, but also contribute directly to the determination of the hadron’s quantum numbers and dynamical properties. This pronounced impact of double-gluon degrees of freedom distinguishes these states from conventional mesons as well as single-gluon hybrid configurations. Moreover, it is theoretically expected that hybrid mesons may accommodate multiple valence gluons, leading to a richer spectrum of exotic hadronic states~\cite{Meyer:2004jc,Kuti:2005xg}. These states play a crucial role in understanding the gluonic binding mechanisms inside hadrons and are also important in the context of possible mixing effects with glueball states, which makes the distinction between glueballs and hybrid mesons nontrivial~\cite{Meyer:2015eta,Klempt:2007cp,Dudek:2013yja}. In particular, lattice QCD studies have demonstrated that double-gluon contributions can have a significant impact on the hybrid meson spectrum and may lead to quantum number assignments different from those of single-gluon hybrids~\cite{HadronSpectrum:2012gic,Bicudo:2018jbb}.

Recently, double-gluon hybrid mesons have been proposed as a new class of exotic hadronic states and have first been investigated in the light-quark sector~\cite{Chen:2021smz,Su:2022fqr,Su:2023jxb}, followed by subsequent studies in the heavy-quark sector~\cite{Tang:2021zti,Su:2023aif,Lian:2024fsb}, where these systems have been systematically extended and studied within the QCD sum rule framework. These investigations indicate that the inclusion of double-gluon degrees of freedom leads to a richer internal structure compared to single-gluon hybrid configurations and significantly modifies the resulting mass spectra and quantum number assignments.

In a recent work~\cite{Esmer:2026jlb}, we have studied light double-gluon hybrid mesons using QCD sum rules by constructing appropriate interpolating currents and evaluating the corresponding two-point correlation functions. That analysis demonstrated that the considered currents provide stable sum rule windows and yield consistent predictions for the masses and current couplings of the light double-gluon hybrid states. Motivated by these findings, it is natural to extend the same framework to the heavy-quark sector.

Motivated by these findings, it is natural to extend the same framework to the heavy-quark sector. In the present work, we generalize the interpolating currents used in Ref.~\cite{Esmer:2026jlb} by replacing the light quark fields with heavy quark fields ($q \rightarrow Q$), and investigate the corresponding double-gluon heavy hybrid meson states within the QCD sum rule approach. In particular, we consider the $\bar{c}GGc$, $\bar{b}GGb$, and $\bar{b}GGc$ systems, and provide systematic predictions for their spectroscopic properties. The spectroscopic analysis is presented in Section~\ref{sec:Hqqbarggmass}, while the numerical analysis and discussion of the results are given in Section~\ref{sec:Numeric}. Finally, Section~\ref{sec:Dis} is devoted to the conclusions of this work.


\section{Spectroscopic Parameters}

\label{sec:Hqqbarggmass}

The masses and current couplings of the heavy double-gluon hybrid states $\bar{c}GGc$, $\bar{b}GGb$, and $\bar{b}GGc$ are extracted in this work within the framework of the two-point QCD sum rule approach. We denote the quark content generically as $\bar{Q}GGQ^{\prime}$, with $(Q,Q^{\prime})=(c,c)$, $(b,b)$, and $(b,c)$. Our analysis begins with the two-point correlation function constructed from the interpolating current $J(x)$. For the spin-0 channels it reads
\begin{equation}
\Pi(q)=i\int d^{4}x\,e^{iqx}\langle 0|\mathcal{T}\{J(x)J^{\dagger}(0)\}|0\rangle ,
\label{eq:CF1}
\end{equation}
while the spin-1 channels, whose currents carry two Lorentz indices, are described by
\begin{equation}
\Pi_{\alpha\beta,\alpha^{\prime}\beta^{\prime}}(q)
=i\int d^{4}x\,e^{iqx}
\langle 0|\mathcal{T}\{J_{\alpha\beta}(x)
J_{\alpha^{\prime}\beta^{\prime}}^{\dagger}(0)\}|0\rangle .
\label{eq:CF2}
\end{equation}

The interpolating currents are obtained by combining a color-octet heavy quark bilinear with a color-octet two-gluon operator into a color singlet, using either the symmetric ($d^{nrs}$) or the antisymmetric ($f^{nrs}$) $\mathrm{SU}(3)$ structure constants~\cite{Su:2023aif}. Among the eight channels $J^{\mathrm{PC}}=0^{\pm\pm}$ and $1^{\pm\pm}$, four currents are
nonvanishing:
\begin{eqnarray}
J_{0^{++}} &=& \bar Q_a \gamma_5 \lambda_n^{ab} Q^{\prime}_b~d^{nrs}~g_s^2 G_r^{\mu\nu} \tilde G_{s,\mu\nu} \, ,
\label{def:0pp}
\\
J_{0^{-+}} &=& \bar Q_a \gamma_5 \lambda_n^{ab} Q^{\prime}_b~d^{nrs}~g_s^2 G_r^{\mu\nu} G_{s,\mu\nu} \, ,
\label{def:0np}
\\
J^{\alpha\beta}_{1^{+-}} &=& \bar Q_a \gamma_5 \lambda_n^{ab} Q^{\prime}_b~f^{nrs}~g_s^2 G_r^{\alpha\mu} \tilde G_{s,\mu}^{\beta} - \{ \alpha \leftrightarrow \beta \} \, ,
\label{def:1pn}
\\
J^{\alpha\beta}_{1^{--}} &=& \bar Q_a \gamma_5 \lambda_n^{ab} Q^{\prime}_b~f^{nrs}~g_s^2 G_r^{\alpha\mu} G_{s,\mu}^{\beta} - \{ \alpha \leftrightarrow \beta \} \, .
\label{def:1nn}
\end{eqnarray}
The currents with $J^{\mathrm{PC}}=0^{+-},\,0^{--},\,1^{++}$, and $1^{-+}$ vanish identically. In the spin-0 cases, the antisymmetric $f^{nrs}$ is contracted with a gluonic structure that is symmetric under $r\leftrightarrow s$. In the spin-1 cases, the symmetric $d^{nrs}$ yields a tensor that is symmetric in $\alpha\leftrightarrow\beta$, which is eliminated by the antisymmetrization. Hence, these four channels are not considered in this work. For the open-flavor $\bar{b}GGc$ states the charge conjugation is not a good quantum number, and the four currents above correspond to $J^{\mathrm{P}}=0^{+},\,0^{-},\,1^{+}$, and $1^{-}$, respectively.

In these expressions, $G^n_{\mu\nu}$ is the gluon field strength tensor, $\tilde G^n_{\mu\nu}=G^{n,\rho\sigma}\epsilon_{\mu\nu\rho\sigma}/2$ is its dual, $\lambda_n$ are the Gell-Mann matrices, $a,b=1,\ldots,3$ and $n,r,s=1,\ldots,8$ are color indices, and $\mu,\nu,\rho,\sigma,\alpha,\beta$ are Lorentz indices. The structure constants satisfy
\begin{equation}
d_{abc}\, d_{abd} = \frac{N^2-4}{N}\,\delta_{cd},\qquad
f_{abc}\, f_{abd} = N\,\delta_{cd},
\label{eq:sym&antisym}
\end{equation}
with $N=3$ for $\mathrm{SU}(3)$.

The masses and current couplings of the heavy double-gluon hybrids are determined by equating two descriptions of the same two-point correlation function. In the first, the correlator is written through the physical properties of the hybrid states; in the second, it is calculated in the operator product expansion (OPE) using the heavy-quark masses, the strong coupling, and the vacuum condensates of increasing mass dimension. Matching the coefficients of the independent Lorentz structures on the two sides leads to the QCD sum rules, whose auxiliary parameters are then fixed according to the standard procedure of the method.

We begin with the hadronic side. Inserting a complete set of states carrying the quantum numbers of the interpolating current into the correlators of Eqs.~\eqref{eq:CF1}and~\eqref{eq:CF2} isolates the contribution of the hybrid state of interest. For illustration, we take the current $J^{\alpha\beta}_{1^{--}}$, which couples to the vector hybrid $H_V$ with quark content $\bar{Q}GGQ^{\prime}$ and $J^{\mathrm{PC}}=1^{--}$ (for
$Q\neq Q^{\prime}$ only $J^{\mathrm{P}}=1^{-}$ is defined). Performing the integral over $x$, the ground-state term is
\begin{eqnarray}
&&\Pi _{\alpha\beta,\alpha^{\prime}\beta^{\prime}}^{\mathrm{Phys}}(q)
=\frac{\langle 0|J^{\alpha\beta}|H_{V}(q,\varepsilon )\rangle
\langle H_{V}(q,\varepsilon )|J_{\alpha^{\prime}\beta^{\prime}}^{\dagger }|0\rangle }
{m_{H_{V}}^{2}-q^{2}}+\ldots,  \notag \\
&&  \label{eq:PhysSide}
\end{eqnarray}
with $m_{H_{V}}$ the mass of the vector hybrid and the ellipses representing
excited and continuum states. The coupling of the current to the hybrid is
encoded in the matrix element
\begin{eqnarray}
\langle 0|J^{\alpha\beta}|H_{V}(q,\varepsilon )\rangle &=& i f_{H_{V}}
\varepsilon^{\alpha\beta\mu\nu} \varepsilon_\mu q_\nu ,
\nonumber \\
\langle H_{V}(q,\varepsilon )|J_{\alpha^{\prime}\beta^{\prime}}^{\dagger }|0\rangle
&=& i f_{H_{V}} \varepsilon^{\alpha^{\prime}\beta^{\prime}\mu^{\prime}\nu^{\prime}}
\varepsilon^*_{\mu^{\prime}} q_{\nu^{\prime}} ,
\label{eq:Mel1}
\end{eqnarray}
in which $f_{H_{V}}$ is the current coupling and $\varepsilon_\mu$ the
polarization vector. Using Eq.~\eqref{eq:Mel1} in Eq.~\eqref{eq:PhysSide} and
summing over polarizations with
\begin{equation}
\sum_{\lambda}\varepsilon^{(\lambda)}_\mu \varepsilon^{*(\lambda)}_{\mu^{\prime}}
= - g_{\mu\mu^{\prime}} + \frac{q_\mu q_{\mu^{\prime}}}{m_{H_V}^2},
\end{equation}
the hadronic representation becomes
\begin{eqnarray}
&&\Pi^{\mathrm{Phys}}_{\alpha\beta,\alpha^{\prime}\beta^{\prime}}(q)
=-\frac{f_{H_V}^2}{m_{H_V}^2-q^2}\Bigg[-q_\beta q_{\beta^{\prime}}g_{\alpha\alpha^{\prime}}
\notag \\
&&+q_\beta q_{\alpha^{\prime}}g_{\alpha\beta^{\prime}}
+q_\alpha q_{\beta^{\prime}}g_{\beta\alpha^{\prime}}
-q_\alpha q_{\alpha^{\prime}}g_{\beta\beta^{\prime}}\notag \\
&&+q^2(g_{\alpha\alpha^{\prime}}g_{\beta\beta^{\prime}}
-g_{\alpha\beta^{\prime}}g_{\beta\alpha^{\prime}})\Bigg]+\cdots .
\label{eq:PhysExpanded}
\end{eqnarray}

Several independent Lorentz structures appear in Eq.~\eqref{eq:PhysExpanded}. For the sum rules we use the one proportional to $g_{\alpha\alpha^{\prime}}g_{\beta\beta^{\prime}}$, which singles out the vector hybrid and allows a stable comparison with the QCD side. We denote its coefficient by $\Pi^{\mathrm{Phys}}(q^{2})$.

The QCD representation is obtained from the same correlation function by contracting the heavy-quark fields in the interpolating currents and expanding the result in the OPE. The time-ordered product of the currents is expressed through local operators of increasing mass dimension, whose vacuum expectation values, the gluon condensates, carry the nonperturbative
information. For the current $J^{\alpha\beta}_{1^{--}}$ we find
\begin{eqnarray}
&&\Pi _{\alpha\beta,\alpha^{\prime}\beta^{\prime}}^{\mathrm{OPE}}(q)=i\int
d^{4}xe^{iqx}{\lambda }_{n}^{ab}{\lambda }_{n^{\prime }}^{a^{\prime }b^{\prime }}f^{nrs}f^{n^{\prime }r^{\prime }s^{\prime }}g_{s}^{4} \notag \\
&&\times \langle 0|(G_r^{\alpha\mu}(x)G_{s,\mu}^\beta(x)-G_r^{\beta\mu}(x)G_{s,\mu}^\alpha(x)) \notag \\
&&\times(G_{s^{\prime },\mu^{\prime }}^{\beta^{\prime }}(0)G_{r^{\prime }}^{\alpha^{\prime }\mu^{\prime }}(0)-G_{s^{\prime },\mu^{\prime }}^{\alpha^{\prime }}(0)G_{r^{\prime }}^{\beta^{\prime }\mu^{\prime }}(0))|0\rangle \notag \\
&&\times \mathrm{Tr}\left[ \gamma _{5}S_{Q^{\prime}}^{bb^{\prime }}(x)\gamma
_{5}S_{Q}^{a^{\prime }a}(-x)\right] ,  \label{eq:QCD}
\end{eqnarray}
where $S_{Q}^{ab}(x)$ is the propagator of the heavy quark $Q=c$ or $b$, and
$S_{Q^{\prime}}^{ab}(x)$ is obtained by replacing $m_{Q}\to m_{Q^{\prime}}$.
Throughout this work we use the expression
\begin{eqnarray}
&&S_{Q}^{ab}(x)=i\int \frac{d^{4}k}{(2\pi )^{4}}e^{-ikx}\Bigg \{\frac{\delta
_{ab}\left( {\slashed k}+m_{Q}\right) }{k^{2}-m_{Q}^{2}}\notag \\
&&-\frac{g_{s}G_{ab}^{\alpha \beta }}{4}\frac{\sigma _{\alpha \beta }\left( {\slashed
k}+m_{Q}\right) +\left( {\slashed k}+m_{Q}\right) \sigma _{\alpha \beta }}{
(k^{2}-m_{Q}^{2})^{2}}  \notag \\
&&+\frac{g_{s}^{2}G^{2}}{12}\delta _{ab}m_{Q}\frac{k^{2}+m_{Q}{\slashed k}}{
(k^{2}-m_{Q}^{2})^{4}}+\frac{g_{s}^{3}G^{3}}{48}\delta _{ab}\frac{\left( {
\slashed k}+m_{Q}\right) }{(k^{2}-m_{Q}^{2})^{6}}  \notag \\
&&\times \left[ {\slashed k}\left( k^{2}-3m_{Q}^{2}\right) +2m_{Q}\left(
2k^{2}-m_{Q}^{2}\right) \right] \left( {\slashed k}+m_{Q}\right) +\cdots
\Bigg \}.  \notag \\
&&  \label{eq:Prop}
\end{eqnarray}
Here $G_{ab}^{\alpha\beta}=G_{n}^{\alpha\beta}t^{n}_{ab}$ with
$t^{n}=\lambda^{n}/2$, and $G^{2}$ and $G^{3}$ denote
$G^{n}_{\alpha\beta}G^{n\,\alpha\beta}$ and
$f^{nml}G^{n}_{\alpha\beta}G^{m\,\beta\gamma}G^{l\,\alpha}_{\ \ \gamma}$,
respectively. 

Equation~\eqref{eq:QCD} contains vacuum averages of products of gluon field-strength tensors, which are evaluated in the OPE at large Euclidean momentum. These averages can be treated in two ways. In the first, their vacuum average is expressed through the gluon condensate, which gives the nonperturbative contribution and is implemented through the
local matrix element
\begin{eqnarray}
\label{eq:Gcond}
&&\langle 0 |G_r^{\alpha\mu}(x)G_{s^{\prime },\mu^{\prime }}^{\beta^{\prime }}(0)|0 \rangle =
\frac{\langle G^2\rangle }{96}\delta^{rs^{\prime }} [g_{\alpha \beta^{\prime }} g_{\mu \mu^{\prime }} \notag \\
&&-g_{\alpha \mu^{\prime }} g_{\beta^{\prime }\mu }].
\end{eqnarray}
Here $\langle G^{2}\rangle\equiv\langle G^{n}_{\alpha\beta}G^{n\,\alpha\beta}\rangle$,
and the factors of $g_s$ are written explicitly where they occur. In the second, the two gluon fields from the currents are contracted with each other by means of the perturbative gluon propagator in the coordinate space, which yields the perturbative contribution,
\begin{eqnarray}
\label{eq:Gprop}
&&\langle 0 |G_r^{\alpha\mu}(x)G_{s^{\prime },\mu^{\prime }}^{\beta^{\prime }}(0)|0 \rangle  =
\frac{\delta^{rs^{\prime }}}{2 \pi^2 x^4} [g_{\mu \mu'}(g_{\alpha \beta^{\prime }}-\frac{4 x_{\alpha} x_{\beta^{\prime }}}{x^2}) \nonumber \\
&& +(\mu, \mu^{\prime }) \leftrightarrow (\alpha, \beta^{\prime })
 -\mu \leftrightarrow \alpha -\mu^{\prime } \leftrightarrow \beta^{\prime }].
\end{eqnarray}
Both treatments are included in the OPE and enter the sum rules together.

Besides these terms, the heavy-quark propagator in Eq.~\eqref{eq:Prop} contains gluon condensates of different dimensions. The dimension-4 and dimension-6 terms, proportional to $\langle g_{s}^{2}G^{2}\rangle$ and $\langle g_{s}^{3}G^{3}\rangle$, appear in the final expressions through the corresponding components of the propagator. The contributions of dimension 8, 10, and 12 are evaluated using the factorization hypothesis for higher-dimensional condensates. This assumption is not exact and is known to be violated for condensates of high dimension~\cite{Ioffe:2005ym}. Nevertheless, we neglect the uncertainties associated with this violation, since the higher-dimensional terms themselves give only small contributions.

The sum rules that determine the mass $m_{H_{V}}$ and the current coupling $f_{H_{V}}$ follow from matching the coefficient of the structure $g_{\alpha\alpha^{\prime}}g_{\beta\beta^{\prime}}$ in the OPE side, $\Pi^{\mathrm{OPE}}(q^{2})$, to the corresponding one in $\Pi^{\mathrm{Phys}}(q^{2})$. After the Borel transformation and the continuum subtraction, we get
\begin{equation}
m_{H_{V}}^{2}=\frac{\Pi ^{\prime }(M^{2},s_{0})}{\Pi (M^{2},s_{0})},  \label{eq:Mass}
\end{equation}
and
\begin{equation}
f_{H_{V}}^{2}=\frac{e^{m_{H_{V}}^{2}/M^{2}}}{m_{H_{V}}^{2}}%
\Pi (M^{2},s_{0}),  \label{eq:Coupl}
\end{equation}
Here, $\Pi^{\prime}(M^{2},s_{0})\equiv d\Pi(M^{2},s_{0})/d(-1/M^{2})$, while $\Pi(M^{2},s_{0})$ is the Borel-transformed and continuum-subtracted OPE amplitude $\Pi^{\mathrm{OPE}}(q^{2})$, which depends on the Borel parameter $M^{2}$ and the continuum threshold $s_{0}$. Eqs.~\eqref{eq:Mass} and \eqref{eq:Coupl} constitute the sum rules for the mass and the current coupling of the vector hybrid state. In evaluating $\Pi(M^{2},s_{0})$, we account for the vacuum condensates up to dimension twelve, and the resulting expression takes the form
\begin{equation}
\Pi (M^{2},s_{0})=\int_{(m_{Q}+m_{Q^{\prime}})^{2}}^{s_0}ds\rho ^{
\mathrm{OPE}}(s)e^{-s/M^2}+\Pi (M^{2}),  \label{eq:CF3}
\end{equation}
where $\rho^{\mathrm{OPE}}(s)$ is the spectral density and $\Pi(M^{2})$ accounts for the terms that are computed directly after the Borel transformation, without a spectral representation. The same procedure applied to the other nonvanishing currents gives the sum rules for the corresponding hybrid states. For the spin-0 channels, no Lorentz structure has to be singled out, since the invariant amplitude coincides with the entire correlation function in Eq.~\eqref{eq:CF1}.

\section{Numerical Results}
\label{sec:Numeric}

The sum rules for the masses and current couplings of the heavy double-gluon hybrids depend on several input parameters. In the numerical computations we use the heavy-quark masses given by the PDG and the gluon condensates:
\begin{eqnarray}
&&m_{c}(\mu )=(1.2729\pm 0.0027)~\mathrm{GeV},\notag \\
&&m_{b}(\mu )=(4.1859\pm 0.0034)~\mathrm{GeV},  \notag \\
&&\langle \alpha _{s}G^{2}/\pi \rangle =(0.012\pm 0.004)~\mathrm{GeV}^{4},
\notag \\
&&\langle g_{s}^{3}G^{3}\rangle =(0.57\pm 0.29)~\mathrm{GeV}^{6}.
\label{eq:GluonCond}
\end{eqnarray}

Here $m_{c}(\mu)$ and $m_{b}(\mu)$ are the $\overline{\mathrm{MS}}$ running masses of the heavy quarks, taken at the scales $\mu=m_{c}$ and $\mu=m_{b}$,
respectively~\cite{ParticleDataGroup:2026mpi}. The gluon condensates were
determined from analyses of hadronic processes~\cite{Shifman:1978bx,Shifman:1978by,Narison:2015nxh}.

The sum rules in Eqs.~\eqref{eq:Mass} and \eqref{eq:Coupl} depend on two auxiliary parameters, the Borel parameter $M^{2}$ and the continuum threshold $s_{0}$. As these parameters are not physical, their working regions are chosen for each channel according to the standard requirements of the method. The ground-state pole should give the dominant part of the correlation function, the OPE should converge with the higher-dimensional terms remaining small, and the extracted masses and current couplings should depend only
weakly on $M^{2}$ within the chosen window. To quantify the pole dominance, we define the pole contribution as
\begin{equation}
\mathrm{PC}=\frac{\Pi (M^{2},s_{0})}{\Pi (M^{2},\infty )},  \label{eq:PC}
\end{equation}
and introduce the ratio
\begin{equation}
R(M^{2})=\frac{\Pi ^{\mathrm{DimN}}(M^{2},s_{0})}{\Pi (M^{2},s_{0})},
\label{eq:Conv}
\end{equation}
where $\Pi ^{\mathrm{DimN}}(M^{2},s_{0})=\sum_{\mathrm{N}=8,10,12}\Pi ^{
\mathrm{DimN}}$ is the sum of the dimension-8, 10, and 12 contributions, which
are proportional to $\langle g_{s}^{2}G^{2}\rangle ^{2}$,
$\langle g_{s}^{2}G^{2}\rangle,  \langle g_{s}^{3}G^{3}\rangle $, and
$\langle g_{s}^{3}G^{3}\rangle ^{2}$, respectively.

We now turn to the vector $\bar{b}GGb$ hybrid state. According to our numerical analysis, for $H_{V}$ the parameter windows
\begin{equation}
M^{2}\in \lbrack 12,14]~\mathrm{GeV}^{2},\ s_{0}\in \lbrack 165,175]~\mathrm{
GeV}^{2},  \label{eq:Wind1}
\end{equation}
\begin{figure}[h]
\includegraphics[width=8.5cm]{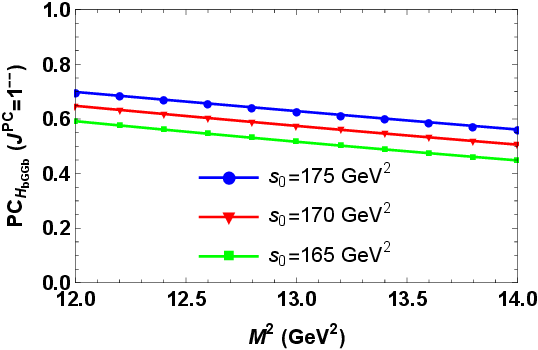}
\caption{Pole contribution $\mathrm{PC}$ as a function of $M^{2}$ at fixed $s_{0}$ for the $\bar{b}GGb$ hybrid meson.}
\label{fig:PC}
\end{figure}
satisfy the constraints imposed by the QCD sum rule analysis. Thus, at $M^{2}=14~\mathrm{GeV}^{2}$ and $M^{2}=12~\mathrm{GeV}^{2}$ on the average in $s_{0}$ the pole contribution is $\mathrm{PC}\simeq 0.51$ and $\mathrm{PC}\simeq 0.65$, respectively. At $M^{2}=12~\mathrm{GeV}^{2}$ the parameter $R(M^{2})$ is negative and less than $0.01$. The dependence of $\mathrm{PC}$ on $M^{2}$ is displayed in Fig.~\ref{fig:PC}. Consequently, for the mass $m_{H_{V}}$ and the current coupling $f_{H_{V}}$ of the $\bar{b}GGb$ hybrid meson $H_{V}$ we obtain
\begin{eqnarray}
m_{H_{V}} &=&(12.02\pm 0.26)~\mathrm{GeV},\ \   \notag \\
f_{H_{V}} &=&(1.54\pm 0.04)~\mathrm{GeV}^{3}.
\end{eqnarray}
The mass $m_{H}$ and the current coupling $f_{H}$ are plotted as functions of $M^{2}$ and $s_{0}$ in Figs.~\ref{fig:Mass} and \ref{fig:Decay constant}, respectively.

The analysis above, carried out for the vector $\bar{b}GGb$ hybrid, is repeated for all the remaining states. The $\bar{c}GGc$ and $\bar{b}GGb$ hybrids are treated in the channels $J^{PC}=0^{++}, 0^{-+}, 1^{+-}$, and $1^{--}$, while the open-flavor $\bar{b}GGc$ states are labeled by $J^{P}=0^{+}, 0^{-}, 1^{+}$, and $1^{-}$. For each of them, the working
regions of the Borel parameter $M^{2}$ and the continuum threshold $s_{0}$, the pole contributions, the convergence ratios $R(M^{2})$, and the resulting masses and current couplings are summarized in Table~\ref{tab:HGG_heavy}. The last column of the table gives the available literature values for
comparison.

\begin{widetext}

\begin{figure}[h!]
\begin{center}
\includegraphics[totalheight=6cm,width=8cm]{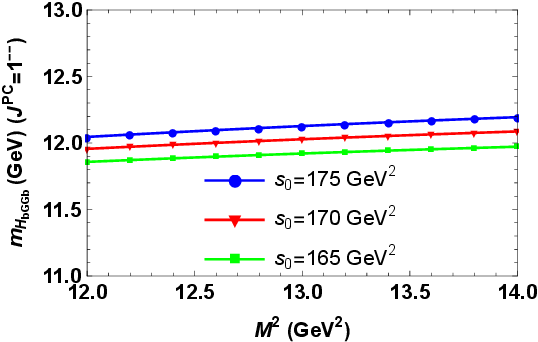}\,\, 
\includegraphics[totalheight=6cm,width=8cm]{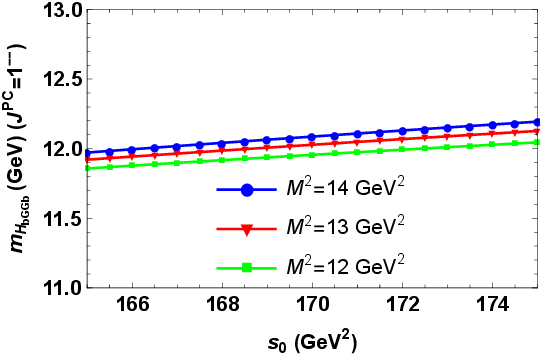}
\end{center}
\caption{ The mass of the $\bar{b}GGb$ hybrid meson plotted against $M^2$ and $s_0$.}
\label{fig:Mass}
\end{figure}

\begin{figure}[h!]
\begin{center}
\includegraphics[totalheight=6cm,width=8cm]{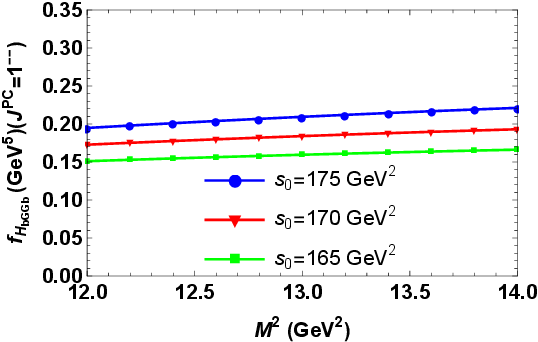}\,\, 
\includegraphics[totalheight=6cm,width=8cm]{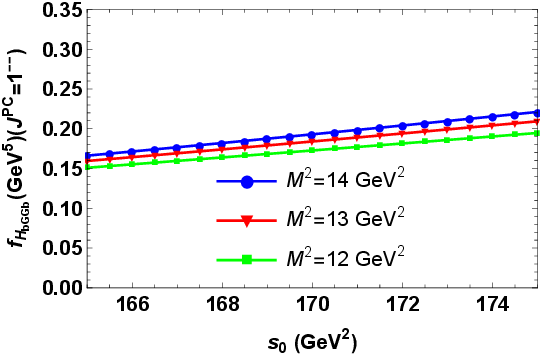}
\end{center}
\caption{The current coupling of the $\bar{b}GGb$ hybrid meson plotted against $M^2$ and $s_0$.}
\label{fig:Decay constant}
\end{figure}

\begin{table}[tbp]
\begin{tabular}{|c|c|c|c|c|c|c|c|c|}
\hline\hline
$J^{\mathrm{PC}}$ & Hybrid & $M^2~(\mathrm{GeV}^2)$ & $s_0~(\mathrm{GeV}^2)$ & $
\mathrm{PC} (\%)$ &$ \left| R(M^{2}) \right| (\%) $& Mass $(\mathrm{GeV})$ & $f (\mathrm{GeV}^3)$
& Mass $(\mathrm{GeV})$ \cite{Su:2023aif} \\ \hline
$0^{++}$ & $\bar{c}GGc$ & $7-9$ & $55-65$ & $46-72$ & $<2.2$ & $6.72\pm 0.29$&$0.47\pm 0.14$&$ 7.29^{+0.33}_{-0.25} $ \\
$0^{-+}$ & $\bar{c}GGc$ & $7-9$ & $55-65$ & $47-73$ & $<5.4$ & $6.52\pm 0.31$&$0.42\pm 0.13$&$ 5.70^{+0.43}_{-0.57} $ \\
$1^{+-}$ & $\bar{c}GGc$ & $7-9$ & $50-60$ & $44-62$ & $<2.3$ & $6.16\pm 0.17$&$0.04\pm 0.01$&$ 6.92^{+0.32}_{-0.12} $ \\
$1^{--}$ & $\bar{c}GGc$ & $7-9$ & $50-60$ & $50-69$ & $<4.4$ & $6.27\pm 0.21$&$0.05\pm 0.01$&$ 5.39^{+0.43}_{-0.56} $ \\ \hline

$0^{++}$ & $\bar{b}GGb$ & $12-14$ & $170-180$ & $50-65$ & $<0.2$ & $12.17\pm 0.27$&$1.97\pm 0.14$&$ - $ \\
$0^{-+}$ & $\bar{b}GGb$ & $12-14$ & $170-180$ & $50-65$ & $<0.2$ & $12.16\pm 0.28$&$1.96\pm 0.14$&$ - $ \\
$1^{+-}$ & $\bar{b}GGb$ & $12-14$ & $165-175$ & $51-65$ & $<0.1$ & $12.02\pm 0.26$&$1.54\pm 0.04$&$ - $ \\
$1^{--}$ & $\bar{b}GGb$ & $12-14$ & $165-175$ & $51-65$ & $<0.1$ & $12.02\pm 0.26$&$1.54\pm 0.04$&$ - $ \\ \hline
$0^{+}$ & $\bar{b}GGc$ & $10-12$ & $110-120$ & $52-71$ & $<1.6$ & $9.57\pm 0.22$&$0.61\pm 0.14$&$ - $ \\
$0^{-}$ & $\bar{b}GGc$ & $10-12$ & $110-120$ & $52-71$ & $<4.4$ & $9.42\pm 0.23$&$0.74\pm 0.16$&$ - $ \\
$1^{+}$ & $\bar{b}GGc$ & $10-12$ & $105-115$ & $54-86$ & $<3.4$ & $9.20\pm 0.22$&$0.13\pm 0.02$&$ - $ \\
$1^{-}$ & $\bar{b}GGc$ & $10-12$ & $105-115$ & $54-72$ & $<5.1$ & $9.58\pm 0.17$&$0.14\pm 0.03$&$ - $ \\ \hline\hline
\end{tabular}
\centering
\caption{QCD sum rule results for the heavy double-gluon hybrid states. Literature values in the last column are available only for $\bar{c}GGc$. The open-flavor $\bar{c}GGb$ states are classified by $J^{P}$.}
\label{tab:HGG_heavy}
\end{table}

\end{widetext}

\section{Conclusions}

\label{sec:Dis} 
We have studied heavy double-gluon hybrid mesons with the quark contents $\bar{c}GGc$, $\bar{b}GGb$, and $\bar{b}GGc$ in the framework of the two-point QCD sum rule method. The interpolating currents combine a color-octet heavy-quark bilinear with a color-octet two-gluon operator. The hidden-flavor states were considered with $J^{PC}=0^{\pm\pm}$ and
$1^{\pm\pm}$, and the open-flavor states with $J^{P}=0^{\pm}$ and $1^{\pm}$. The currents with $J^{PC}=0^{+-}, 0^{--}, 1^{++}$, and $1^{-+}$ vanish identically, so the sum rules were derived for the four remaining channels. The OPE was computed including condensates up to dimension twelve, and the Borel parameter and continuum threshold were fixed by requiring a sizable pole contribution and a convergent OPE. In this way we extracted the masses
and current couplings of all these states. The quoted uncertainties come from the variations of the Borel parameter and continuum threshold within their working windows and from the errors of the input parameters.

For $\bar{c}GGc$, as far as we are aware, the only available prediction constructed with the same interpolating currents and quantum numbers is that of Ref.~\cite{Su:2023aif}, and we therefore compare our results with it. Our masses differ from those values by roughly $8$--$16\%$. The $0^{++}$ and $0^{-+}$ results lie close to the edges of the quoted uncertainty ranges, whereas the $1^{+-}$ and $1^{--}$ masses differ more clearly. These differences may be traced to the different input parameters and working windows in the two analyses and deserve further study. For the $\bar{b}GGb$ and $\bar{b}GGc$ hybrids, to the best of our knowledge, no earlier predictions are available, so the present results extend the spectroscopy of double-gluon hybrids to the bottomonium-like and open-flavor sectors and
cannot yet be compared with other works. The masses grow with the heavy-quark content, with the open-flavor $\bar{b}GGc$ states lying between the $\bar{c}GGc$ and $\bar{b}GGb$ ones. The obtained masses and current couplings can serve as input for future theoretical studies, for example of the decay widths and of these states in other frameworks, and as guidance for
experimental searches for heavy hybrids with two valence gluons.

\section*{ACKNOWLEDGMENTS}

The work of G.~Daylan~Esmer, H.~Sundu, B.~Barsbay and S.~T\"{u}rkmen is supported in part by the Scientific and Technological Research Council of T\"{u}rkiye (TUBITAK) under Grant No.~125F556.


\end{document}